\documentclass[aps,prl,twocolumn,showpacs,superscriptaddress,groupedaddress,nofootinbib]{revtex4}  
\usepackage{bm}
\usepackage{mathrsfs}
\usepackage{amsmath}
\usepackage{amssymb}
\usepackage{graphicx}
\usepackage{amsfonts}
\usepackage{cancel}
\usepackage{amsthm}
\usepackage{color}
\usepackage{dcolumn}
\usepackage{txfonts}
\usepackage{subfigure}
\begin{document}


\title{Imaginary geometric phases of quantum trajectories}

\author{ Fan Yang }
\affiliation{Department of Physics, The Chinese University of Hong Kong, Shatin, N.T., Hong Kong, China}

\author{ Ren-Bao Liu }
\email{rbliu@phy.cuhk.edu.hk}
\affiliation{Department of Physics, The Chinese University of Hong Kong, Shatin, N.T., Hong Kong, China}
\affiliation{Centre for Quantum Coherence, The Chinese University of Hong Kong, Shatin, N.T., Hong Kong, China}
\affiliation{Institute of Theoretical Physics, The Chinese University of Hong Kong, Shatin, N.T., Hong Kong, China}



\begin{abstract}

A quantum object can accumulate a geometric phase when it is driven along a trajectory in a parameterized state space with non-trivial gauge structures. Inherent to quantum evolutions, a system can not only accumulate a quantum phase but may also experience dephasing, or quantum diffusion. Here we show that the diffusion of quantum trajectories can also be of geometric nature as characterized by the imaginary part of the geometric phase. Such an imaginary geometric phase results from the interference of geometric phase dependent fluctuations around the quantum trajectory. As a specific example, we study the quantum trajectories of the optically excited electron-hole pairs, driven by an elliptically polarized terahertz field, in a material with non-zero Berry curvature near the energy band extremes. While the real part of the geometric phase leads to the Faraday rotation of the linearly polarized light that excites the electron-hole pair, the imaginary part manifests itself as the polarization ellipticity of the terahertz sidebands. This discovery of geometric quantum diffusion extends the concept of geometric phases.

\end{abstract}
\maketitle

When a discrete quantum eigenstate is adiabatically driven in the parameterized state space, in addition to the familiar dynamical phase, the state can acquire a geometric phase which depends on the gauge structure of the quantum system. In particular, the geometric phase accumulated along a cyclic evolution is the famous Berry phase, which is a gauge-invariant physical quantity~\cite{Berry,phase}. The geometric phase has played essential roles in many fields of physics, such as the Aharonov-Bohm effect~\cite{AB_phase}, quantum Hall effect~\cite{Avron,TKNN,QNiu,Prange,QNiu2}, anomalous Hall effect ~\cite{QNiu2,AH1,AH2,AH3,AH4} and topological insulators~\cite{SCZhang,Kane}.

In contrast to a discrete eigenstate, a wavepacket in a continuum is a superposition of infinitely many eigenstates. For example, an electron in the energy bands of a semiconductor is described by a wavepacket as a superposition of the Bloch states. When the wavepacket is driven adiabatically along a trajectory in the parameter space, it can pick up a geometric phase similar to a discrete state~\cite{Chang_Niu1995,Chang_Niu1996,QNiu2}. In addition, however, the wavepacket during the evolution will experience quantum diffusion (or dephasing) due to interference between different phase factors associated with different energy eigenstates that form the wavepacket. In this Letter, we show that the quantum diffusion can also have a geometric origin, due to the geometric part of the quantum phase of each eigenstate. This geometric quantum diffusion is characterized by an imaginary geometric phase, which is determined by the geometry of the quantum evolution in the parameterized state space.

\begin{figure}
\begin{center}
\includegraphics[width=\columnwidth]{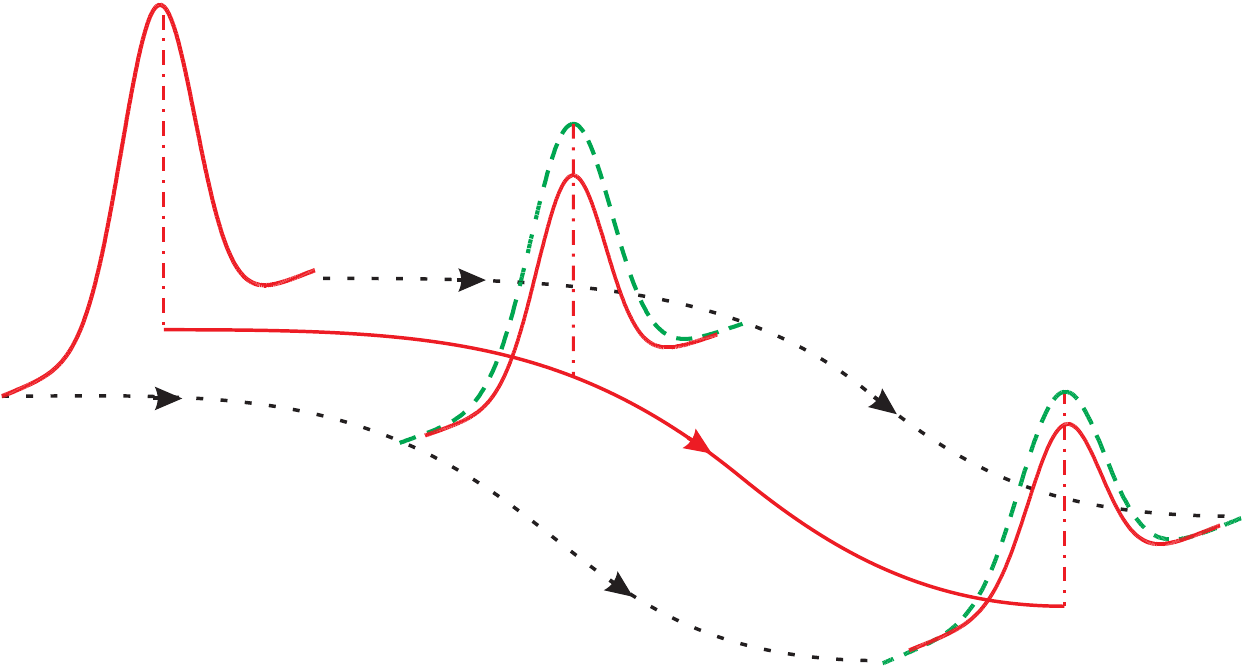}
\end{center}
\caption{(color online). Schematics of the quantum trajectories of a wavepacket in the parameter space. The green dashed (red solid) Gaussian curves represent the diffusion of the wavepacket without (with) the geometric diffusion included. The red solid arrow represents the quantum trajectory that satisfies the stationary phase condition.} \label{fig_intro}
\end{figure}

We consider the path integral form of the adiabatic evolution of the wavepacket (see Fig. \ref{fig_intro})
\begin{equation}
\left\langle \mathbf R \right| G\left(t_i , t_f \right) \left| W_i \right\rangle=\int {c\left( {\mathbf R_i } \right)e^{iS\left( {\mathbf R_{i \to f} } \right)} } \left\langle {\mathbf R} \mathrel{\left | {\vphantom {\mathbf R {\mathbf R_f }}} \right. \kern-\nulldelimiterspace} {{\mathbf R_f }} \right\rangle {\mathcal D}\mathbf R_{i \to f},
\end{equation}
where $\left| \mathbf R \right\rangle$ denotes an eigenstate with parameter $\mathbf R$, $\left| W_i \right\rangle = \int {c\left( {\mathbf R_i } \right)\left|\mathbf R_i\right\rangle } d\mathbf R_{i}$ is the initial wavepacket at time $t_i$, $G$ is the propagator from $t_i$ to $t_f$, $\mathbf R_{i \to f}$ denotes a path from $\mathbf R_i$ to $\mathbf R_f$ and $S\left( {\mathbf R_{i \to f} } \right)$ gives the action (i.e., phase) of the evolution along this path. In adiabatic evolution, the action $S\left( {\mathbf R_{i \to f} } \right)$ can be decomposed into the dynamical part $S_D$ and the geometric part $S_G$. To explore the geometric phase effects, it is helpful to consider the semiclassical approximation, in which the summation of the phase factors along all possible paths is dominated by the orbits that satisfy the stationary phase condition $\delta S[\mathbf R_{cl}] = 0$, called quantum trajectories~\cite{Path_integral,HHG_QT}. Then the propagator is determined by the semiclassical actions of the quantum trajectories plus fluctuations nearby:
\begin{equation}
\left\langle \mathbf R \right| G \left| W_i \right\rangle \approx \sum\limits_{\mathbf R_{cl}} e^{iS^{\left(cl\right)}}\int {c\left( {\mathbf R_i } \right)e^{i \delta^2 S^{\left(cl\right)}} } \left\langle {\mathbf R} \mathrel{\left | {\vphantom {\mathbf R {\mathbf R_f }}} \right. \kern-\nulldelimiterspace} {{\mathbf R_f }} \right\rangle {\mathcal D}\mathbf R_{i \to f},
\end{equation}
where
\begin{equation}
\delta^2 S^{\left( {cl} \right)}  = \frac{1}{2}\int_{t_i }^{t_f } {\left. {\frac{{\delta^2 S\left[\mathbf R \right]}}{{\delta \mathbf R\left( {t_1 } \right) \delta \mathbf R\left( {t_2 } \right)}}} \right|} _{\mathbf R_{cl} } \mathbf q\left( {t_1 } \right) \mathbf q\left( {t_2 } \right)dt_1 dt_2
\end{equation}
gives the quantum fluctuation around the quantum trajectories with $\mathbf q = \mathbf R - \mathbf R_{cl}$. Note that $S^{\left(cl\right)}=S_G[\mathbf R_{cl}]+S_D[\mathbf R_{cl}]$ naturally contains a geometric action $S_G$, which depends only on the geometry of the quantum trajectories.

In the general case (e.g., in quantum tunneling~\cite{leggett_RMP}), there may be no real orbit obeying the classical equation of motion. Then we have to invoke analytic continuation of the classical mechanics to the complex plane. As a result, the geometric phase of the quantum trajectory also becomes complex with a nonzero imaginary part of the action. Since the imaginary part of the action $\Im S^{\left(cl\right)}$ describes the quantum diffusion of the wavepacket due to the quantum interference, $\Im S_G \left[\mathbf R_{cl}\right]$ represents the geometric part of the quantum diffusion. This geometric quantum diffusion results from the interference of the different geometric phase factors associated with a bunch of paths (quantum fluctuations) near the quantum trajectory.

\begin{figure}
\begin{center}
\includegraphics[width=\columnwidth]{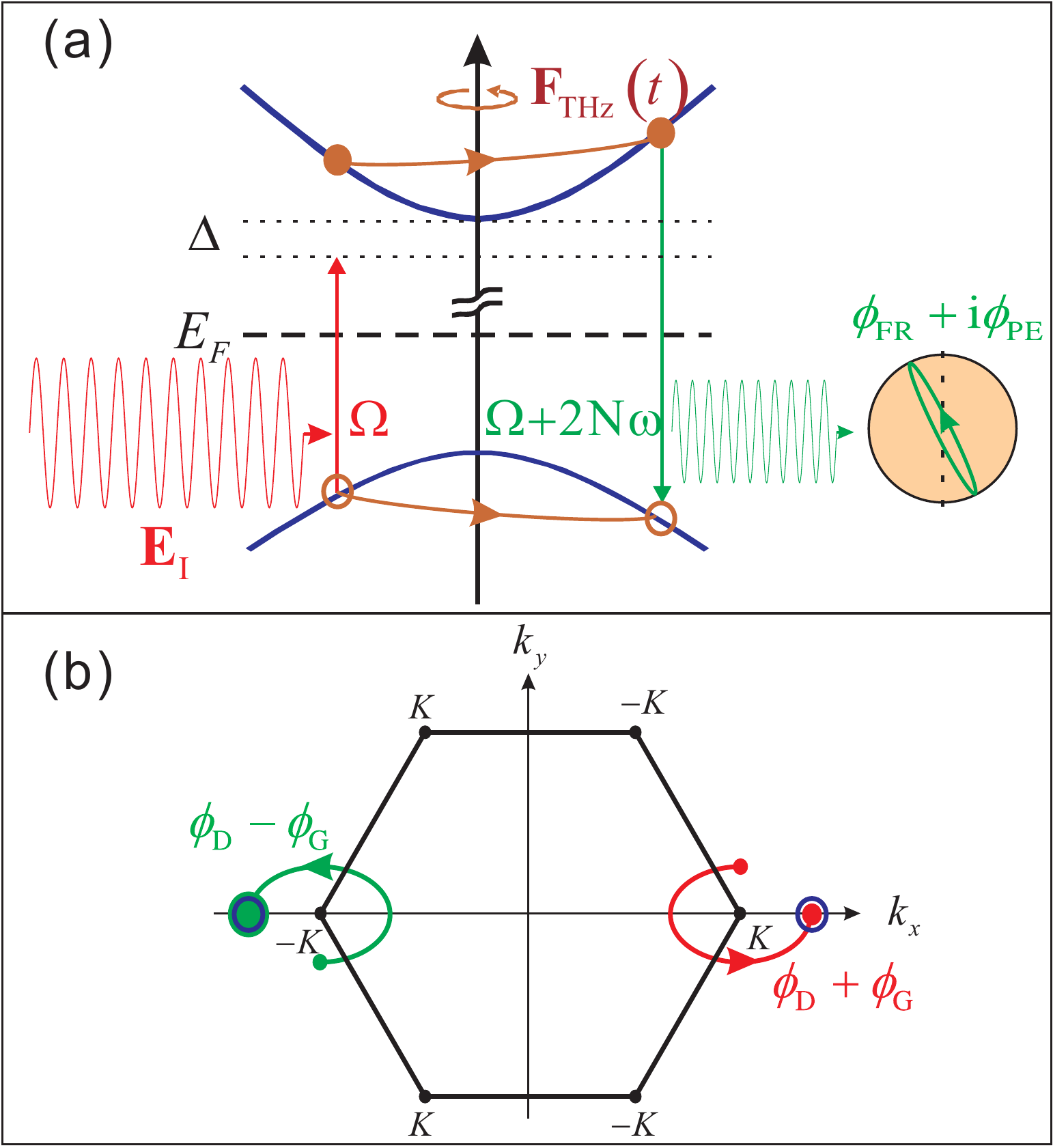}
\end{center}
\caption{(color online). Schematics of HSG and quantum trajectories in monolayer MoS$_2$. (a) An electron-hole pair excited by a linearly polarized optical laser ${\mathbf E}_{\text{I}}e^{-i\Omega t}$ is driven along an elliptical quantum trajectory by the THz field, acquiring a kinetic energy, and recombines with emission of a sideband photon at frequency $\Omega+2N\omega$. The sideband has a polarization ellipticity $\phi_{PE}$ and Faraday rotation $\phi_{FR}$, given by the imaginary and real parts of the geometric phase, respectively. (b) Interference of the quantum trajectories that contribute to the $N$th order sideband. The blue open circles represents the wavepackets at $\pm K$ valleys with dynamical diffusion only, while the filled circles are the wavepackets including the geometric diffusion. } \label{Schem}
\end{figure}

To give a specific example of the geometric quantum diffusion, we consider the quantum trajectories of an optically excited electron-hole pair, driven by a terahertz (THz) field in a semiconductor. The electron-hole pair, after excitation by a weak optical laser of frequency $\Omega$, is driven into oscillations by an intense THz laser of frequency $\omega$. The electron-hole pair subsequently acquires a kinetic energy, and recombines at a later time to emit photons at sideband frequencies $\Omega+2N\omega$, with $N$ being an integer [Fig.~\ref{Schem}(a)]. This so-called high-order THz sideband generation (HSG) has been theoretically studied and recently experimentally observed~\cite{HSG_RBL,Zaks_Liu,HSG_Hunter}. In HSG the electron-hole wavepacket evolution is well approximated by a small number of quantum trajectories plus the quantum fluctuations around them. When the THz field is elliptically polarized, the quantum trajectories become curved. Our previous study shows that geometric phases will be accumulated along these trajectories if the energy bands of the semiconductors (such as monolayer MoS$_2$ and bilayer graphene) have non-vanishing Berry curvatures~\cite{YF_MoS2,YF_bilayer}. The geometric phases result in an observable effect in time-reversal symmetric materials: The optical emission at integer multiples of the THz period after the excitation by a linearly polarized laser pulse has a Faraday rotation (FR) equal to the Berry phase accumulated along the quantum trajectory~\cite{YF_MoS2,YF_bilayer}. Therefore HSG in such materials provides an ideal platform for studying the geometric quantum diffusion (i.e., the imaginary geometric phase).

In this Letter, we demonstrate that the imaginary geometric phases have observable effects on HSG in time-reversal symmetric materials with non-vanishing Berry curvatures. The geometric and dynamical phases can be separated by their different behaviors under the time-reversal transformation: the geometric phase is reversed while the dynamical one is unchanged. By interference between time-reversal related quantum trajectories, we can observe the real part of the geometric phase as an FR of the sideband emission~\cite{YF_MoS2,YF_bilayer}, and the imaginary part as the polarization ellipticity (PE) of the sidebands.

The elliptically polarized THz field can be written as
\begin{equation}\label{Ft}
{\mathbf F}\left( t \right) = F \left( \cos \theta \cos \left(\omega t\right), \sin \theta \sin \left(\omega t\right), 0\right),
\end{equation}
where $F$ is the field strength. Under this field, the electron and hole will evolve along an elliptic path in the $\mathbf k$-space:
\begin{align} \label{path}
\tilde {\mathbf k}\left( t \right) = \left( {k_x  - k_0\cos\theta \sin \left(\omega t\right),k_y  + k_0\sin\theta \cos \left(\omega t\right)},k_z \right),
\end{align}
where $k_0  = {eF}/\omega$. The driving by the THz field is adiabatic in the sense that the THz field has a frequency much lower than the band gap of the material and hence induces no interband transition. The excitation by a weak linearly polarized optical laser is described by the interaction Hamiltonian $\hat H_{\text{I}}=-\hat{{\mathbf P}}\cdot {\mathbf E}_{\text{I}}e^{-i\Omega t}+\text{h.c.}$. Here $\hat{{\mathbf P}}= \int {d{\mathbf k}} \hat e_{\mu, {\mathbf k}}^\dag \hat h_{\nu, -{\mathbf k}}^\dag {\mathbf{d}}_{\mu \nu,{\mathbf k}}$ is the interband polarization operator,
where $\hat e_{\mu, \mathbf k}$ ($\hat h_{\nu,\mathbf k}$) annihilates an electron (hole) with momentum $\mathbf k$ and spin or valley index $\mu$ ($\nu$), and the interband dipole moment ${{\mathbf d}}_{\mu \nu ,{{\mathbf k}}}$ is~\cite{Blount,Chang_Niu1996}
\begin{equation}
{{\mathbf d}}_{\mu \nu ,{{\mathbf k}}}=\frac{{e\left\langle {+,\mu , {\mathbf k}} \right|i\nabla _{\mathbf k} H\left(\mathbf k\right)  \left| {-,\nu , {\mathbf k}} \right\rangle }}{{E^+_{\mathbf k}  - E^-_{\mathbf k} }},
\end{equation}
with $+$ and $-$ denoting the conduction and valence bands, respectively, and $E_{{\mathbf k}}^{\pm}$ the band energy. We assume that the semiconductor is initially in the ``vacuum" state $\left|G\right\rangle$ with empty conduction bands and filled valence bands. Then the linear optical response is~\cite{YF_MoS2}
\begin{align}
{\mathbf P}\left( t \right) = &\sum_{\mu} i\int_{ - \infty }^t {dt'} \int {d {\mathbf k}} {{\mathbf d}}_{\mu\mu,\tilde {{\mathbf k}} \left( t \right)}^* {{\mathbf d}}_{\mu\mu,\tilde {{\mathbf k}} \left( {t'} \right)}  \cdot {{\mathbf E}}_{\text{I}} \notag\\
& e^{ - i\int_{t'}^t {\varepsilon _{\tilde {{\mathbf k}}\left( \tau \right)} d\tau + i \int_{t'}^t { \left[{{\mathscr {A}}_{\tilde {{\mathbf k}}  \left( \tau \right)}  }\right]_{\mu\mu} \cdot d\tilde {{\mathbf k}} \left(\tau\right)}  - i\Omega t'} } , \label{AbelQT}
\end{align}
where $\varepsilon _{\tilde {{\mathbf k}}}  = E^{+}_{\tilde {{\mathbf k}}}  - E^{-}_{\tilde {{\mathbf k}}}$ is the energy of the electron-hole pair, and ${\mathscr{A}}_{\tilde{\mathbf k}}={\mathscr{A}}^+_{\tilde{\mathbf k}}- {\mathscr{A}}^-_{\tilde{\mathbf k}}$ is the combined Berry connection of the electron-hole pair. Here for the sake of simplicity we have assumed that the Berry connection is Abelian (i.e., the THz field does not mix (near) degenerate bands of different spin or valley indices). The generalization to non-Abelian case is possible.

We take monolayer MoS$_2$ as a model system. This material has two time-reversal related valleys $\pm K$ at the corners of the 2D hexagonal Brillouin zone [Fig.~\ref{Schem}(b)], where the strong spin-orbit coupling causes a spin splitting of about 160 meV at the valence band top~\cite{MOS2}. We assume that the optical laser is near-resonant with the transitions between the band edges of the conduction band and the highest valence band, and therefore neglect the transitions from the lower valence bands. The energy bands near the band edge can be effectively described by the Hamiltonian~\cite{MOS2}
\begin{equation}\label{Hamiltonian_MoS2}
H\left( {\mathbf k} \right) = A\left( {\xi k_x \sigma _x  + k_y \sigma _y } \right) + M\sigma _z,
\end{equation}
where $A=3.51 \ {\rm{eV}} \cdot \AA$, the band gap $2M=1.9 \ {\rm{eV}}$, $\xi=\pm 1$ denotes the $\pm K$ valley, and $\mathbf k$ is measured from the respective Dirac points at valleys $\pm K$~\cite{MOS2,MOS2_Zeng,MOS2_Mak}. The energy spectrum is $\varepsilon_{{\mathbf k}}=2\sqrt{M^2+A^2 k^2}\approx 2M+A^2k^2/M$ with two-fold valley degeneracy. The Berry connection and dipole moment at $\pm K$ valleys (labelled by the pseudo-spin $\Uparrow/\Downarrow$) satisfy the time-reversal relations
\begin{subequations}
\begin{equation}
\left({\mathscr{A}}_{{\mathbf k}}\right) _{\Uparrow  \Uparrow} = \left({\mathscr{A}}_{ - {\mathbf k}}\right)_{\Downarrow \Downarrow}^*=-\left({\mathscr{A}}_{ {\mathbf k}}\right)_{\Downarrow \Downarrow}=\frac{A^2} {2M^2}\left( { k_y {\textbf{e}}_x } -k_x {\textbf{e}}_y \right), \label{TRrelation1}
\end{equation}
\begin{equation}
{{\mathbf d}}_{\Uparrow  \Uparrow,{\mathbf k}} = {{\mathbf d}}_{\Downarrow  \Downarrow, - {\mathbf k}}^*=d_{cv}\frac{ {\textbf{e}}_x  -i{\textbf{e}}_y }{\sqrt{2}}, \label{TRrelation2}
\end{equation}
\end{subequations}
where $d_{cv}=i\frac{eA}{\sqrt{2}M}$. Note that the optical selection rules are such that the interband transition at valley $+K$ ($-K$) is coupled exclusively with the $\sigma+$ ($\sigma-$) polarized light.

The effects of the geometric phases on the HSG in monolayer MoS$_2$ or similar materials can be intuitively understood. A laser with linear polarization $\sigma^{+}+\sigma^{-}$ causes equal transitions in $+K$ and $-K$ valleys at time $t-\tau$ and creates an electron-hole pair in the superposition state $\left| \Uparrow \right\rangle  + \left| \Downarrow \right\rangle$. After the driving by the THz field, the quantum trajectories at the two valleys obtain the same dynamical phase $\phi_D$ but opposite geometric phases $\pm\phi_G$ [Fig.~\ref{Schem}(b)], so the superposition becomes ${e^{i\phi _G} \left|  \Uparrow \right\rangle  + e^{ - i\phi_G } \left| \Downarrow \right\rangle } $. The emission by the recombination of the electron-hole pair at time $t$ will have the polarization $ e^{i\phi_G} \sigma^+ +e^{-i\phi_G} \sigma^-$. The real part of $\phi_G$ induces a phase shift between the two circular polarizations and hence an FR $\Re\phi_G$~\cite{YF_MoS2,YF_bilayer}, while the imaginary part induces an amplitude difference between the two circular polarizations and hence a PE $\Im\phi_G$ (assumed $\ll 1$) [Fig.~\ref{Schem}(a)].

\begin{figure}
\begin{center}
\includegraphics[width=\columnwidth]{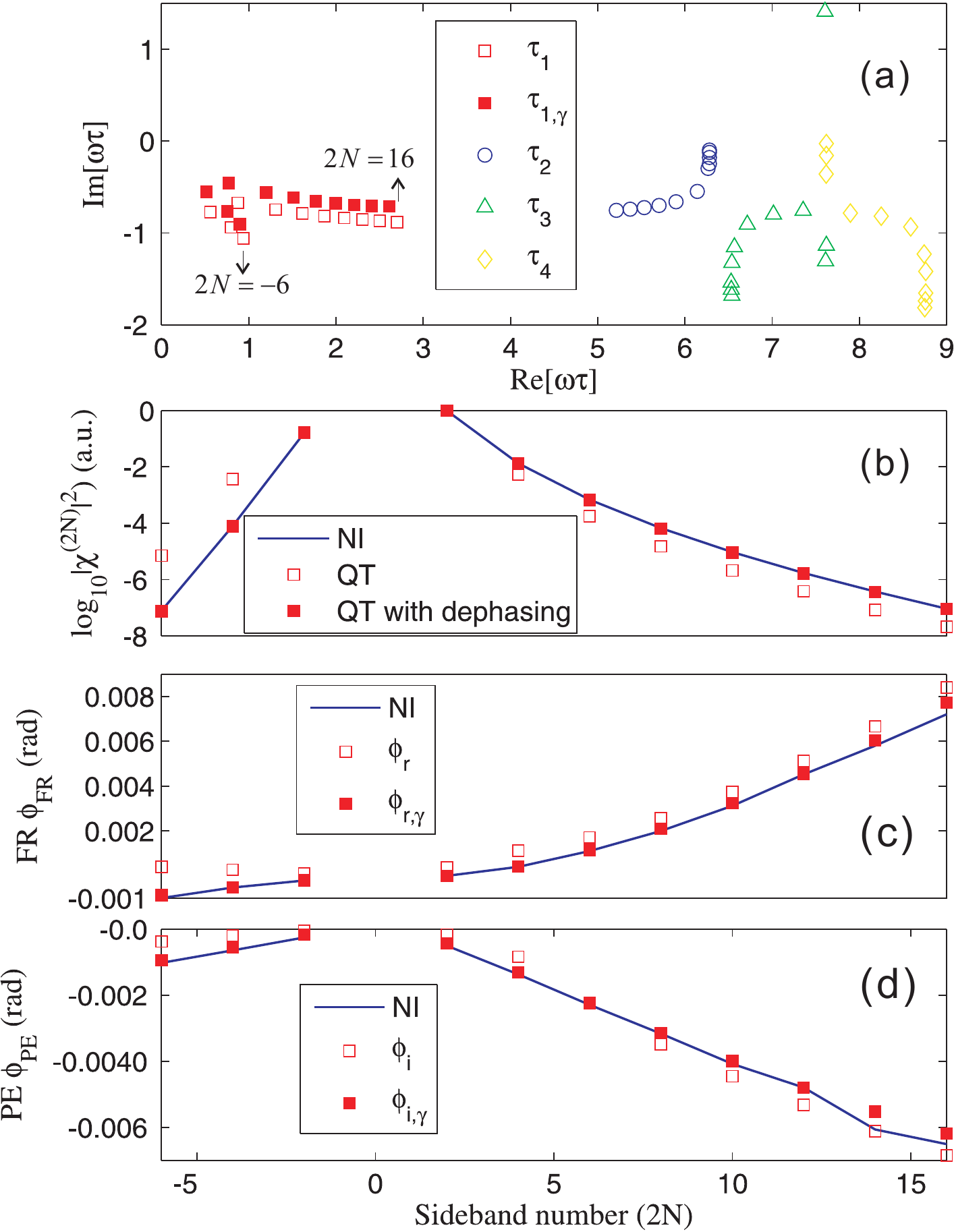}
\end{center}
\caption{(color online). Faraday rotation and polarization ellipticity of THz sidebands in monolayer MoS$_2$. (a) shows the stationary phase points $\tau_n$ for $S_{cl}^{\left(2N\right)}$ with the sideband order $2N=-6 \to 16$. Only the first four stationary phase points ($n=1\to 4$) are given for each sideband. The filled squares are the first ($n=1$) stationary phase points $\tau_{1,\gamma}$ that includes the dephasing effect [solutions of Eq. (\ref{St_tau_gamma})]. (b) shows the relative intensities of the sidebands, while (c) and (d) give the corresponding FR and PE of them. In (b-d), the lines are obtained by numerical integration of Eq. (\ref{suscept_2N}), and the open and filled squares show the quantum trajectory (QT) results obtained using the dominant stationary phase points $(t_{1},\tau_{1})$ [the first physical solution to Eq. (\ref{St_tau})] and $(t_{1,\gamma},\tau_{1,\gamma})$ [the first physical solution to Eq. (\ref{St_tau_gamma})], respectively. } \label{final_results}
\end{figure}

The susceptibility of the $N$th sideband (at frequency $\Omega+2N\omega$) for the $\sigma_{\pm}$-polarized optical field is
\begin{equation}\label{suscept_2N}
\chi^{\left(2N\right)}_{ \pm \pm}  = i \left|d_{cv}\right|^2 \int dt \int_{0}^{\infty}d\tau \int d{\mathbf{k}} {e^ {iS^{\left(2N\right)}\left( {\mathbf{k},t,\tau} \right)\pm i\phi_G \left( {\mathbf{k},t,\tau} \right)}},
\end{equation}
where $\tau=t-t'$ denotes the evolution time between the excitation and the emission,
\begin{equation}\label{geometric_phase}
\phi_G \left( {\mathbf{k},t,\tau} \right)=\frac{eA^2}{2M^2}\int_{t-\tau}^t {\left[ {{\tilde k_x \left( {t_1} \right)} {\mathbf{e}}_y  - {\tilde k_y\left( {t_1} \right)}{\mathbf{e}}_x } \right] \cdot {\mathbf{F}}\left( {t_1} \right)dt_1}
\end{equation}
is the geometric phase, and the action
\begin{equation}
S^{\left(2N\right)} \left( {\mathbf{k},t,\tau} \right) =  - \int_{t-\tau}^t {  \frac{\tilde {\mathbf{k}}^2 \left( {t_1} \right)}{2m^*} dt_1}  -\Delta\tau + 2N\omega t,\label{Mos2_action}
\end{equation}
with $m^* = M/\left(2A^2\right)$ being the reduced effective mass of the electron-hole pair and $\Delta=2M-\Omega$ the optical laser detuning. Since $\phi_G/S^{\left(2N\right)} \sim \omega/\left(2M\right) \ll 1$, the quantum trajectories are determined by the stationary phase conditions for the dynamical phase, i.e.,
\begin{equation}
\frac{\partial S^{\left(2N\right)} \left( {\mathbf{k},t,\tau} \right)}{\partial {\mathbf k} } =  - \int_{t-\tau}^t {  \frac{\tilde {\mathbf{k}} \left( {t_1} \right)}{m^*} dt_1} =0.\label{Sk}
\end{equation}
Note that ${\tilde {\mathbf{k}} \left( {t_1} \right)}/{m^*}$ is the semiclassical velocity of the electron-hole pair. Hence equation (\ref{Sk}) means the electron accelerated by the THz field returns to the hole after $\tau$ for recombination. By Eq. (\ref{geometric_phase}), the geometric phase accumulated along the quantum trajectory determined by Eq. (\ref{Sk}) is
\begin{equation}\label{geometric_tau}
\phi_G \left(\tau\right) =  -\phi_B \frac{\omega \tau }{2\pi}\left[1-{\text{sinc}}^2\left( \frac{\omega\tau}{2} \right)\right],
\end{equation}
with $\phi_B=\pi \sin\theta \cos\theta k_0^2 A^2/M^2$ being the Berry phase an electron-hole pair acquires in a full THz period. By the saddle point approximation (which approximates the electron-hole evolution by quantum trajectories plus quantum fluctuations around them), Eq. (\ref{suscept_2N}) becomes~\cite{XTXie,HSG_Yan}
\begin{equation}\label{suscept_QT}
\chi^{\left(2N\right)}_{ \pm \pm} = \sum\limits_n {i \left|d_{cv}\right|^2  \frac{2\pi m^{*}{e^ {iS^{\left(2N\right)}_{cl}\left( t_n,\tau_n \right)}}}{i\tau_n+0^{+}} \sqrt {\frac{{\left( {2\pi i} \right)^2 }}{{\det \left[ \partial ^2 S_{cl}^{\left( {2N} \right)} \right]}}} {e^ {\pm i\phi_{G} \left( \tau_n\right)}} },
\end{equation}
where $\left(t_n,\tau_n\right)$ is the stationary phase point satisfying
\begin{equation}
\frac{\partial S_{cl}^{\left(2N\right)} \left( {t,\tau} \right)}{\partial {t} }=\frac{\partial S_{cl}^{\left(2N\right)} \left( {t,\tau} \right)}{\partial \tau }=0,\label{St_tau}
\end{equation}
with $S_{cl}^{\left(2N\right)} \left( {t,\tau} \right)=S^{\left(2N\right)} \left( {\mathbf k_{cl},t,\tau} \right)$ and $\mathbf k_{cl}$ being the solution of Eq. (\ref{Sk}). Generally, there are no real solutions to equation (\ref{St_tau}), i.e., $t_n$ and $\tau_n$ are complex numbers [Fig.~\ref{final_results}(a)]. They hence determine a complex quantum trajectory and lead to an imaginary part of the geometric phase $\phi_{G}\left(\tau_n\right)$. In particular, if one of the stationary phase points dominates, say $(t_1,\tau_1)$, we have $\chi^{\left(2N\right)}_{\pm\pm} \propto {e^ {\pm i\phi_{G} \left( \tau_1\right)}}$. If the optical laser is polarized along ${\mathbf e}_x = \left(\sigma^+ + \sigma^-\right)/\sqrt 2$, the polarization of the $N$th sideband is
\begin{equation}
\frac{\sigma^+ e^{i\phi_r - \phi_i  }  + \sigma^- e^{-i\phi_r + \phi_i  }}{\sqrt 2} = {\mathbf e}_{\parallel}\cosh \phi_i - i{\mathbf e}_{\bot}\sinh \phi_i, \label{FR_PE}
\end{equation}
where ${\mathbf e}_{\parallel}= {\mathbf e}_x \cos \phi_r  - {\mathbf e}_y \sin \phi_r$ and ${\mathbf e}_{\bot}=
{\mathbf e}_x \sin \phi_r  + {\mathbf e}_y \cos \phi_r$ with $\phi_r$ and $\phi_i$ being the real and imaginary parts of the geometric phase $\phi_{G}(\tau_1)$, respectively. That gives the FR $\phi_r$ and PE $\phi_i$ (for $\phi_i \ll 1$).

In order to confirm the validity of the quantum trajectory method, we calculate the susceptibilities by direct numerical integration of Eq. (\ref{suscept_2N}), and compare the FR and PE of the sidebands with the real and imaginary parts of the geometric phases at the stationary phase points. In the calculation, the THz field is set such that $\omega=2$ meV, $F_{\text{THz}} = 10$ kV/cm and $\theta=\pi/6$, and the laser is tuned below the band gap by $\Delta=2\omega$. To describe the scattering effect in real materials, we include a phenomenological dephasing term of the electron-hole pair $e^{-\gamma \tau}$ in the integration of Eq. (\ref{suscept_2N}), with $\gamma=3$ meV. Such dephasing can be due to phonon scattering, relaxation of the electron-hole pair to bound exciton states and so on. For the sake of simplicity, we neglect the Coulomb interaction between the electron and hole, which is justified since the exciton binding energy (100s of meV~\cite{MOS2_binding}) is much larger than $\omega$ and $\Delta$ and hence the exciton bound states are far off-resonant from the optical excitation.

The comparison is shown in Fig.~\ref{final_results}. The results of the sideband strength in Fig.~\ref{final_results}(b) suggest that the electron-hole pair evolution is well approximated by the dominant trajectory with $n=1$. The numerically calculated FR and PE of the sidebands are almost equal to the real and imaginary parts of the geometric phases $\phi_{G}$ accumulated along the first quantum trajectory [determined by the first physical solution $(t_1,\tau_1)$ of Eq. (\ref{St_tau})], respectively. The discrepancy is, on the one hand, due to the contribution from other trajectories, and on the other hand, due to the changing of the stationary phase points by the inclusion of the dephasing term:
\begin{equation}\label{Scl_gamma}
S^{\left(2N\right)}_{cl,\gamma} \left( {t,\tau } \right)= S^{\left(2N\right)}_{cl} \left( {t,\tau } \right)+i\gamma\tau.
\end{equation}
The modified stationary phase points $(t_{1,\gamma},\tau_{1,\gamma})$ satisfy
\begin{equation}
\frac{\partial S_{cl,\gamma}^{\left(2N\right)} \left( {t,\tau} \right)}{\partial {t} }=\frac{\partial S_{cl,\gamma}^{\left(2N\right)} \left( {t,\tau} \right)}{\partial \tau }=0. \label{St_tau_gamma}
\end{equation}
$\tau_{1,\gamma}$ and their corresponding geometric phases $\phi_{r/i,\gamma}$ are also shown in Fig.~\ref{final_results}, which indeed agree better with the numerical results.

We observe in Fig.~\ref{final_results} that both the FR and the PE increases almost linearly with the sideband order $2N>0$. This can be understood as follows. The electron-hole pair need to go along a longer trajectory [i.e. larger $\tau$ as shown in Fig.~\ref{final_results}(a)] in the $\mathbf k$-space to acquire a higher kinetic energy $2N\omega-\Delta$, which in turn leads to a larger geometric phase of the wavepacket [see equation (\ref{geometric_tau})].


In summary, we have shown that the geometric phase of a wavepacket along a quantum trajectory can have both real
and imaginary parts. The imaginary part characterizes quantum diffusion of the wavepacket which only depends on the geometry of the quantum trajectory. As an example, we showed that while the real part of the geometric phase leads to a Faraday rotation of the THz sideband emission in a time-reversal symmetric semiconductor, the imaginary part manifests itself as the polarization ellipticity of the sideband. This finding extends the concept of the geometric phase to the complex plane, which may lead to a wealth of new physics.

\begin{acknowledgements}
This work is supported by Hong Kong RGC/GRF 401011 and the CUHK Focused Investments Scheme.
\end{acknowledgements}


\newcommand{\noopsort}[1]{} \newcommand{\printfirst}[2]{#1}
  \newcommand{\singleletter}[1]{#1} \newcommand{\switchargs}[2]{#2#1}

\end{document}